\documentclass[prb,twocolumn,showpacs,aps]{revtex4}
\usepackage{graphicx}
\usepackage{bm}
\usepackage{amssymb}
\usepackage{amsmath}

\input{tcilatex}

\begin{document}

\title{A Phenomenological Theory of The Pseudogap State}
\author{Kai-Yu Yang$^{1}$, T. M. Rice$^{1,2}$ and Fu-Chun Zhang$^{1}$}
\affiliation{$^{1}$ Centre of Theoretical and Computational Physics and Department of
Physics, The University of Hong Kong, Hong Kong\\
$^{2}$ Institut f\"{u}r Theoretische Physik, ETH Zurich, CH-8093 Z\"{u}rich,
Switzerland}
\date{\today }
\pacs{74.20.Mn, 74.25.Jb, 79.60.-i}

\begin{abstract}
An ansatz is proposed for the coherent part of the single particle Green's
function in a doped resonant valence bond (RVB) state by, analogy with the
form derived by Konik and coworkers for a doped spin liquid formed by an
array of 2-leg Hubbard ladders near half-filling. The parameters of the RVB
state are taken from the renormalized mean field theory of Zhang and
coworkers for underdoped cuprates. The ansatz shows good agreement with
recent angle resolved photoemission (ARPES) on underdoped cuprates and
resolves an apparent disagreement with the Luttinger Sum Rule. The
transition in the normal state from a doped RVB spin liquid to a standard
Landau Fermi liquid, that occurs in the renormalized mean field theory,
appears as a quantum critical point characterized by a change in the
analytic form of the Green's function. A d-wave superconducting dome
surrounding this quantum critical point is introduced phenomenologically.
Results are also presented for the Drude weight and tunneling density of
states as functions of the hole density.
\end{abstract}

\maketitle

\section{Introduction}

The cuprate superconductors have attracted enormous interest not just
because of their high temperature unconventional superconductivity but also
because of their highly anomalous properties in the normal phase (for a
review see T. Timusk and B. Statt \cite{T. Timusk}). These show strong
deviations from standard Landau Fermi liquid behavior particularly in the
underdoped region where the deviations are most evident in the pseudogap (or
spin gap) phase. Among the most spectacular of these deviations is the
observation of a Fermi surface in photoemission (ARPES) experiments that
consists, not of a closed contour, but only of 4 disconnected arcs centered
on the Brillouin zone diagonals. \cite{arcs, K. M. Shen} Pseudogaps in the
single particle spectrum truncate the Fermi surface at the saddle points. In
addition the density of charge carriers is determined not by the conduction
electron density but by the hole density doped into the stoichiometric
parent Mott insulator. \cite{T. Timusk} These facts indicate that the
pseudogap phase should be viewed as an anomalous precursor to the
stoichiometric Mott insulator.

Very soon after the discovery of the high temperature superconductors and
before these anomalous pseudogap properties were measured, Anderson proposed
that the key to understanding these unique materials lies in what he called
Resonant Valence Bond (RVB) behavior. \cite{RVB} He proposed a description
based on lightly hole doped spin liquid of spin singlets. Rather than
forming a fixed array of singlets, strong quantum fluctuations among the
antiferromagnetically coupled $S=1/2$ spins lead to a superposition of
singlet configurations, i.e. the bond singlets resonate between many
configurations. This elegant concept explains many key features of the
pseudogap phase as emphasized in the recent review by Anderson and
coworkers. \cite{vanillar} But the strong onsite correlations among the
electrons in this precursor to a Mott insulator are difficult to treat
analytically. This has hindered the development of complete theory for the
RVB phase and the cuprates in general.

In this recent review\cite{vanillar}, Anderson and coworkers point out that
the early renormalized mean field theory (RMF) introduced by Zhang and
coworkers\cite{RMF} for the RVB phase predicted many of its key features
using simple Gutzwiller renormalization factors to describe the strong
correlations. In the intervening years much progress has been made on a
first principle treatment of these correlations based on gauge theories (for
a review see Lee, Nagaosa and Wen \cite{Lee-Nagaosa-Wen}). The functional
renormalization group (RG) approach developed by Honerkamp and coworkers has
also given new insights starting from weak coupling. \cite{Honerkamp}
However, we still lack even a consistent phenomenological description of
this anomalous pseudogap phase that ties together key features analogous to
the Landau theory for standard Fermi liquids. The purpose of this paper is
to take a step in this direction. To this end we introduce an ansatz for the
form of the self-energy and thereby for the form of the single particle
Green's function. In this we are guided by the recent theory by Konik, Rice
and Tsvelik (KRT) for the form of the Green's function in a doped spin
liquid consisting of an array of 2-leg Hubbard ladders coupled by long range
inter-ladder hopping. \cite{KRT} By a careful choice of the form of this
hopping, KRT could justify a random phase approximation (RPA) to obtain the
two dimensional Green's function from the single ladder form. They showed
that for a light doping away from one electron/site a novel form of the
Green's function, $G(\boldsymbol{k},\omega )$ resulted. In particular they
showed that the behavior of $G(\boldsymbol{k},0)$ which enters the Luttinger
Sum Rule (LSR) \cite{Luttinger}, was quite distinct from that of a normal
Landau Fermi liquid. The LSR relates the area enclosed by the contours where 
$G(\boldsymbol{k},0)$ changes sign from positive to negative to the total
electron density. In a Landau Fermi liquid, sign changes in $G(\boldsymbol{k}%
,0)$ occur only through an infinity in $G(\boldsymbol{k},0)$ at a closed
Fermi surface. In the doped spin liquid, KRT found that the sign changes
occur not only through infinities on Fermi pockets but also through zeroes
lying on a separate surface which they call the Luttinger surface of zeroes.
The Fermi pockets enclose an area determined by the hole density while the
Luttinger surface encloses an area given by the density of one electron per
site in the parent stoichiometric ladders. This behavior followed from the
form of the self energy in the coherent part of the single ladder Green's
function and its modification through the inter-ladder hopping processes.
The single ladder Green's function was derived by Essler and Konik only in
the weak coupling limit. \cite{Essler} However, the fact that weak and
strong coupling are continuously connected in ladders and that the form is
qualitatively similar to that obtained numerically by Troyer et al in strong
coupling $t-J$ ladders \cite{Troyer}, leads us to believe that this general
form is characteristic of doped spin liquid and is not restricted to weak
coupling.

In section 2 we introduce our ansatz for the form of the self energy for a
two dimensional lightly doped RVB spin liquid. Our ansatz is based on a
generalization of the KRT form for the doped spin liquid formed in an array
of 2-leg Hubbard ladders. \cite{KRT} It also contains a Luttinger surface of
zeroes which encloses a commensurate density in addition to hole Fermi
surface pockets. In this case the Luttinger surface coincides with the
so-called umklapp surface introduced by Honerkamp and coworkers \cite%
{Honerkamp} as the surface where umklapp scattering processes appear to open
up a charge gap in the weak coupling RG flow.

It is also interesting to note that this Green's function is closely related
to the form recently proposed by Tai-Kai Ng \cite{Tai-Kai Ng}. He started
from the strong coupling limit and introduced spin-charge separation by
factorizing single electron operators into a product of spinon and holon
operators. Following earlier work by Wen and Lee \cite{Wen and Lee} he
introduced a phenomenological attraction between spinon and holon which
leads to a binding between spinon and holon. He obtained a form for the
Green's function which also contains a coherent quasiparticle pole and has
close similarities to our ansatz.

In section 3, we analyze the consequences of our ansatz for a variety of
properties. First we obtain the hole density dependence of our psuedogap and
other key parameters from the RMF results of Zhang et al.\cite{RMF} Based on
these choices we obtain estimates of the hole density dependence of many
observables, e.g. the Fermi velocity $v_{F}$ and wavevector $\boldsymbol{k}%
_{F}$ along the Brillouin zone (nodal) direction, the minimum gap contour
near the saddle point (antinodal direction), the tunneling density of state
(DOS), the Drude weight, and the shape and form of the hole Fermi pockets.
From the phenomenological form of the Green's function we obtain the
quasiparticle dispersion and spectral weights that characterize the coherent
part of the single electron Green's function. A key point in our
phenomenology is that there is a critical hole concentration above which the
spin liquid anomalous self-energy vanishes. As consequence there is a form
of quantum critical point (QCP) which separates two topologically distinct
forms for the Green's function. Below the critical hole density, $G(%
\boldsymbol{k},0)$ is characterized by coexisting Luttinger surfaces of
zeroes and hole pocket Fermi surface of infinities while above the critical
concentration $G(\boldsymbol{k},0)$ displays only a closed Fermi surface of
infinities as usual in a Landau theory. Many features of this QCP resemble
those deduced by Loram and coworkers from an analysis of a variety of
experiments \cite{Loram}.

In section 4, we examine how this Green's function is modified when the
system enters a d-wave superconducting state. This again is introduced
phenomenologically and no attempt is made here to derive the parameters of
the superconductivity although we do offer a number of reasons why the hole
pockets should have a d-wave superconducting instability.

Lastly section 5 is devoted to further discussions and conclusions.

\section{Ansatz for the self energy in a doped RVB spin liquid}

A doped RVB spin liquid has many properties which are quite distinct from
the standard Landau Fermi liquid which follows from treating the
interactions in perturbation theory. Moreover these anomalous properties
cannot be ascribed to a broken symmetry or to the appearance of a new order
parameter. As a result it is a challenge to construct a consistent theory
even on the phenomenological level of a doped RVB spin liquid. We start with
a brief review of the recent work by KRT \cite{KRT} who could obtain the
form of the single particle Green's function $G(\boldsymbol{k},\omega )$ in
a doped spin liquid. In particular, they examined the form of the LSR which
applies to the zero frequency Green's function $G(\boldsymbol{k},0)$ in the
doped spin liquid. The key point about the LSR as emphasized in the famous
textbook by Abrikosov, Gorkov and Dzyaloshinskii (AGD) \cite{AGD} and more
recently by Tsvelik \cite{Tsvelik-QFT}, is that derivation of the LSR is
very general and is not limited to perturbation theory. The LSR relates the
total electron density, $\rho $, to the area in the $\boldsymbol{k}$-space
where $G(\boldsymbol{k},0)>0$. In two dimensions it takes the form%
\begin{equation}
\rho =\frac{2}{(2\pi )^{2}}\int_{G(\boldsymbol{k},0)>0}d^{2}\boldsymbol{k}
\label{LSR}
\end{equation}%
An important point that these authors emphasized is that the sign change
from positive to negative values of $G(\boldsymbol{k},0)$ is not restricted
to an infinity in $G(\boldsymbol{k},0)$ such as occurs at the Fermi surface
of a Landau Fermi liquid. It can also occur through a zero in $G(\boldsymbol{%
k},0)$ as, for example, in the case in the BCS theory of superconductivity.
KRT considered a doped spin liquid consisting of an array of two-leg Hubbard
ladders. At half-filling in a single ladder the Fermi surface consists of
four points without interactions, but it is completely truncated when the
repulsive interactions which lead to both spin and charge gaps, are
introduced. All spin and charge correlation functions are strictly short
range so that this system is a true spin liquid.

An explicit form for the single particle Green's function has been derived
by Konik and coworkers in the limit of weak repulsion. \cite{Essler, Konick}
Around each of the four Fermi points $G(k,\omega )$ can be split into
coherent and incoherent parts, 
\begin{equation}
G_{a}^{L}(k,\omega )=\frac{z_{a}(\omega +\epsilon _{a}(k))}{\omega
^{2}-\epsilon _{a}^{2}(k)-\Delta ^{2}}+G_{inc}  \label{Green_total}
\end{equation}%
Here $\epsilon _{a}(k)$ is the bare dispersion near the corresponding Fermi
wavevector $k_{Fa}$, $z_{a}$ is quasiparticle weight $\sim 1$, and $\Delta $
the single particle gap. Note while the coherent part has a form similar to
the diagonal Green's function in BCS theory, there is no off-diagonal
component of the Green's function in this case. This general form is also
compatible with the numerical results of Troyer et al for the strong
coupling limit \cite{Troyer}. $G_{a}^{L}(k,0)$ changes sign from positive to
negative through a zero in $G_{a}^{L}(k,0)$ which occurs when $\epsilon
_{a}(k)$ passes through zero as $k$ crosses $k_{Fa}$, and the area with $%
G_{a}^{L}(k,0)>0$ is unchanged by the interactions, satisfying the LSR.

Starting from this result, KRT could derive $G_{a}^{L}(\boldsymbol{k},\omega
)$ for an array of 2-leg Hubbard ladders coupled by a particular form of
long range inter-ladder hopping chosen so that it could be treated using a
random phase approximation (RPA) 
\begin{equation}
G_{a}^{RPA}(\boldsymbol{k},\omega )=1/\{G_{a}^{L}(k,\omega )^{-1}-t_{\bot
}(k_{\bot })\}  \label{2leg}
\end{equation}%
where $\boldsymbol{k}=(k,k_{\bot })$. They showed that this form, for values
of $t_{\bot }$ large enough, leads to the appearance of electron and hole
pockets distinct from the lines of zeroes in $G_{a}^{RPA}(\boldsymbol{k},0)$%
. The zero contours remain as lines at $k=k_{Fa}$, independent of the
transverse component of the wavevector, $k_{\bot }$. In the presence of hole
doping, the hole pocket expands and the electron pocket shrinks. In this
doped spin liquid $G_{a}^{RPA}(\boldsymbol{k},0)$ has sign changes both
through infinities along the Fermi pockets and through zeroes located on the
lines $k=k_{Fa}$. Note although the zeroes lines appear at incommensurate
wavevectors in general and not at the Brillouin zone as in a standard Bloch
insulator, the total area they enclose is commensurate and equals 1 electron
/site, irrespective of the inter-ladder hopping strength. The final form
obtained by KRT for the coherent part is 
\begin{equation}
G_{a}^{RPA}(\boldsymbol{k},\omega )=\frac{z_{a}}{\omega -\epsilon
_{a}(k)-t_{\bot }(k_{\bot })-\Delta ^{2}/(\omega +\epsilon _{a}(k))}
\label{2leg2}
\end{equation}%
This can be interpreted as a ladder self energy $\Sigma _{L}(k,\omega
)=\Delta ^{2}/(\omega +\epsilon _{a}(k))$, where $\epsilon _{a}(k)=0$ at the 
$k$-points where the gap opens up in the parent insulating two-leg Hubbard
ladder array.

In this work we consider doping a two-dimensional resonant valence bond
insulator. We start from the RMF for such an insulator. This approximation
treats the effect of strong correlations through renormalization factors
calculated by a description that goes back to early work by Gutzwiller \cite%
{Gutzwiller}. The renormalized $t-J$ Hamiltonian takes the form of an
effective Hamiltonian \cite{RMF}%
\begin{equation}
H_{eff}=g_{t}T+g_{s}J\sum_{\left\langle i,j\right\rangle }\boldsymbol{S}%
_{i}\cdot \boldsymbol{S}_{j}  \label{RMT}
\end{equation}%
with the kinetic, $T$, and spin energy terms modified by factors $g_{t}$ and 
$g_{s}$ respectively 
\begin{eqnarray}
g_{t} &=&\frac{2x}{1+x},  \notag \\
g_{s} &=&\frac{4}{(1+x)^{2}}  \label{factor}
\end{eqnarray}%
for a hole doping of $x$ per site. At half-filling $g_{t}=0$ leaving only
the magnetic energy. The RVB ansatz factorizes the spin energy introducing
both Fock exchange, $\chi _{i,j}=\left\langle c_{i,\sigma }^{\dagger
}c_{j,\sigma }\right\rangle $ and pairing, $\Delta _{i,j}=\left\langle
c_{i,\sigma }c_{j,\sigma }\right\rangle $ expectation values. The
factorization procedure is not unique but the spin quasiparticle dispersion
that results is unique, $E_{\boldsymbol{k}}=(3g_{s}J/8)(\cos ^{2}k_{x}+\cos
^{2}k_{y})^{1/2}$. Upon hole doping $g_{t}>0$, and coherent quasiparticle
poles with a small weight $g_{t}$ appear in the single particle Green's
function. By analogy with the KRT form for the doped spin liquid discussed
above, we make the following ansatz for the coherent part of $G(\boldsymbol{k%
},\omega )$ in a doped RVB spin liquid 
\begin{equation}
G^{RVB}(\boldsymbol{k},\omega )=\frac{g_{t}}{\omega -\xi (\boldsymbol{k}%
)-\Delta _{R}^{2}/(\omega +\xi _{0}(\boldsymbol{k}))}+G_{inc}  \label{PG}
\end{equation}%
where $\boldsymbol{k}=(k_{x},k_{y}),$%
\begin{eqnarray}
\xi _{0}(\boldsymbol{k}) &=&-2t(x)(\cos k_{x}+\cos k_{y})  \notag \\
\Delta _{R}(\boldsymbol{k}) &=&\Delta _{0}(x)(\cos k_{x}-\cos k_{y})  \notag
\\
\xi (\boldsymbol{k}) &=&\xi _{0}(\boldsymbol{k})-4t^{\prime }(x)\cos
k_{x}\cos k_{y}  \notag \\
&&-2t^{\prime \prime }(x)(\cos 2k_{x}+\cos 2k_{y})-\mu _{p}
\label{parameter2}
\end{eqnarray}%
Eq.\ref{PG} is analogy to Eq.\ref{2leg2} for the coupled ladder system, and $%
\xi (\boldsymbol{k})-\xi _{0}(\boldsymbol{k})$ is analogy to $t_{\bot
}(k_{\bot })$ in Eq.\ref{2leg2}. In the renormalized dispersion we include
hopping terms out to 3$^{\text{rd}}$ nearest neighbor with coefficients 
\begin{eqnarray}
t(x) &=&g_{t}(x)t_{0}+\frac{3}{8}g_{s}(x)J\chi ,  \notag \\
t^{\prime }(x) &=&g_{t}(x)t_{0}^{\prime },  \notag \\
t^{\prime \prime }(x) &=&g_{t}(x)t_{0}^{\prime \prime }  \label{parameter}
\end{eqnarray}%
The RVB gap magnitude function $\Delta _{0}(x)$ is also taken from the RMF
theory \cite{RMF}. The parameter $\mu _{p}$ represents a shift of the energy
band so that the chemical potential is always the zero of the energy. We
determine $\mu _{p}$ from the LSR on the total electron density.

First we consider the limit of zero doping, $x\rightarrow 0$. In this case $%
g_{t}(x)\rightarrow 0$ and the quasiparticle dispersion reduces to the
spinon dispersion and the quasiparticles have the vanishing weight in this
limit in the single particle Green's function, $G(\boldsymbol{k},\omega )$.
At small but finite $x$ the zero frequency Green's function $G^{RVB}(%
\boldsymbol{k},0)$ that enters the LSR has lines of zeroes when $\xi _{0}(%
\boldsymbol{k})(=-2t(x)(\cos k_{x}+\cos k_{y}))=0$. The Luttinger contour of
zeroes in $G^{RVB}(\boldsymbol{k},0)$ then consists of straight lines
connecting the points $(\pm \pi ,0)$ and $(0,\pm \pi )$. This Luttinger
contour coincides with the antiferromagnetic (AF) Brillouin zone. But more
relevantly it coincides with the umklapp surface which appears in functional
RG calculations on the weak coupling 2D $t-t^{\prime }-U$ Hubbard model.
Honerkamp \cite{Honerkamp}, Laeuchli and coworkers \cite{Laeuchli} found
that umklapp scattering processes in both particle-hole and
particle-particle channel which connect points on this surface grew strongly
at low energies and temperatures leading them to propose that an energy gap
would open up on this surface below a critical scale. Further because of a
close analogy to the behavior found in the case of a half-filled 2-leg
Hubbard ladder, they proposed that this gap was not driven by long range
order, but rather was a sign that a RVB spin liquid with short range order
would form below a critical scale. Thus our ansatz for the Green's function
is fully consistent with these proposals.

A second feature that follows from our ansatz for $G^{RVB}(\boldsymbol{k}%
,\omega )$ is the appearance of hole pockets at finite hole doping. These
will be reviewed in detail in the next section. The hole pockets define
Fermi surfaces where $G^{RVB}(\boldsymbol{k},0)$ changes sign through
infinities and contain a total area equal the hole density. The LSR is
satisfied since the area with $G^{RVB}(\boldsymbol{k},0)>0$ is bounded by
the Luttinger surface which contains 1 el/site, minus four hole pockets
which have a total area related to the hole density.

The phenomenological form $G^{RVB}(\boldsymbol{k},0)$ for a hole doped RVB
spin liquid can be straightforwardly generalized to a d-wave superconducting
state%
\begin{eqnarray}
&&G_{coh}^{S}(\boldsymbol{k},\omega )  \notag \\
&=&g_{t}/[\omega -\xi (\boldsymbol{k})-\Sigma _{R}(\boldsymbol{k},\omega ) 
\notag \\
&&-\left\vert \Delta _{S}(\boldsymbol{k})\right\vert ^{2}/(\omega +\xi (%
\boldsymbol{k})+\Sigma _{R}(\boldsymbol{k},-\omega ))]  \label{SC}
\end{eqnarray}
where $\Sigma _{R}(\boldsymbol{k},\omega )$ is the RVB spin liquid self
energy from Eq.[\ref{PG}] 
\begin{equation}
\Sigma _{R}(\boldsymbol{k},\omega )=\left\vert \Delta _{R}(\boldsymbol{k}%
)\right\vert ^{2}/(\omega +\xi _{0}(\boldsymbol{k}))  \label{SC_selfenergy}
\end{equation}%
and $\Delta _{S}(\boldsymbol{k})$ is the d-wave superconducting gap
function. To analyze the LSR on $G_{coh}^{S}(\boldsymbol{k},0)$ we note
first that $\Sigma _{R}(\boldsymbol{k},0)\rightarrow \infty $ on the surface
where $\xi _{0}(\boldsymbol{k})=0$. Thus $G_{coh}^{S}(\boldsymbol{k},0)$
continues to have a Luttinger surface of zeroes on the umklapp surface as in
the normal RVB spin liquid. However, there is now an additional set of
Luttinger surfaces defined by the contour that satisfy 
\begin{equation*}
\xi (\boldsymbol{k})+\Sigma _{R}(\boldsymbol{k},0)=0
\end{equation*}%
But these are just the Fermi surface of the four hole pockets in the normal
phase which have now been converted to a Luttinger surface of zeroes in the
superconducting state. Thus the form Eq.[\ref{SC}] continues to satisfy the
LSR.

Lastly we remark that along the Brillouin zone diagonals both $\Delta _{R}(%
\boldsymbol{k})$ and $\Delta _{S}(\boldsymbol{k})$ vanish in our
phenomenological form. As a result exactly along these directions there is
only a single quasiparticle pole which crosses the Fermi energy at a Fermi
wavevector determined by $\xi (\boldsymbol{k})|_{\boldsymbol{k}%
=(k_{F},k_{F})}=0$.

Our phenomenological form for $G^{RVB}$ in the normal and superconducting
phases are similar but not identical to several other recent proposals. It
is very close to the form Tai-Kai Ng \cite{Tai-Kai Ng} derived based on
spin-charge separation but with an added phenomenological attraction between
spinon and holon which leads to binding and therefore to quasiparticle poles
and a self energy form similar to Eq.[\ref{SC_selfenergy}]. He, however,
restricted his analysis to the case of only nearest neighbor hopping.
Earlier Norman and coworkers from an analysis of ARPES data on BSCCO samples
around optimal doping introduced a d-wave superconducting self energy
modified to included both normal state single particle scattering and pair
scattering. \cite{M. R. Norman} Recently Honerkamp \cite{HonerkampPC} has
speculated on an adaptation of the form introduced by Norman et al to
describe a truncation of the Fermi surface through the opening of energy gap
in the antinodal regions near $(\pm \pi ,0)$, $(0,\pm \pi )$ as a
phenomenological description of the RVB spin liquid.

\section{Electronic Properties of the Normal Pseudogap Phase}

In this section we discuss the electronic properties that follow from our
phenomenological form for the Green's function and compare these to
experiments on the normal pseudogap state.

\begin{figure}[tbp]
\includegraphics[width=9.0cm,height=7.0cm]{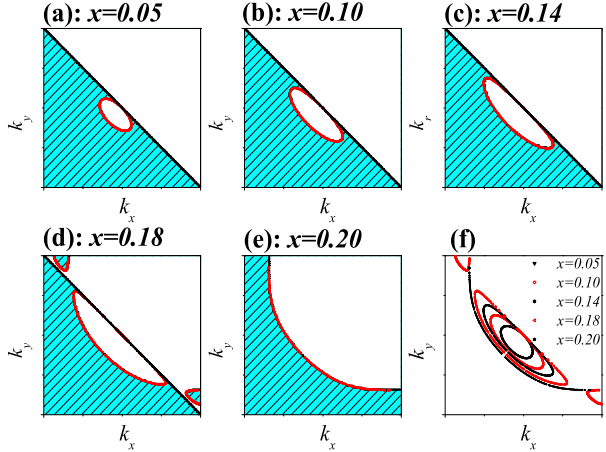}
\caption{(Color online) Contours on which $G(\boldsymbol{k},0)$ changes sign
at various hole concentrations $x$ are shown in (a)-(e). In the shaded area $%
G(\boldsymbol{k},0)>0$, satisfying the Luttinger Sum Rule. In the normal
pseudogap phase, the line connecting $(\protect\pi ,0)-(0,\protect\pi )$ is
the Luttinger surface of zeroes and the pockets in the thick line represent
the infinities of $G(\boldsymbol{k},0)$. The values of the parameters used
here are given in Fig.\protect\ref{parameter_fig}. The evolution of the
contours of infinities in $G(\boldsymbol{k},0)$ is illustrated in (f).}
\label{LSR_fig}
\end{figure}

\begin{figure}[tbp]
\includegraphics[width=9.0cm,height=10.0cm]{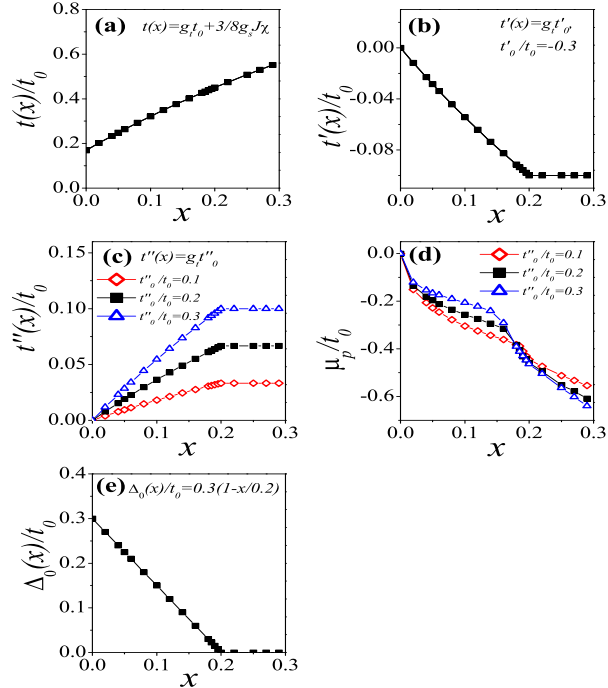}
\caption{(Color online) The values of the parameters $t(x)$, $t^{\prime }(x)$%
, $t^{\prime \prime }(x)$, $\protect\mu _{p}$ and $\Delta (x)$ that enter in
Eq.[\protect\ref{parameter2},\protect\ref{parameter}] used\ in the present
calculations. $(\protect\chi =0.338$, $J/t_{0}=1/3)$ Results presented in
the text are for a choice of $t_{0}^{\prime }/t_{0}=-0.3$, $t_{0}^{\prime
\prime }/t_{0}=0.2$, estimated from the calculated band structure of $%
Ca_{2}CuO_{2}Cl_{2}$. \protect\cite{Mattheiss}}
\label{parameter_fig}
\end{figure}

We begin with the LSR shown in Eq.[\ref{LSR}]. In Fig.\ref{LSR_fig}(a-e) we
show that the contours on which $G^{RVB}(\boldsymbol{k},\omega =0)$, defined
in Eq.[\ref{PG}], changes sign. We chose values for hopping parameters in
Eq.[\ref{parameter}], appropriate to $Ca_{2-x}Na_{x}CuO_{2}Cl_{2}$. The
underlying band structure parameters were obtained by a tight-binding fit to
the antibonding $3d_{x^{2}-y^{2}}-2p_{x(y)}$ band calculated by Mattheiss 
\cite{Mattheiss}. These were then renormalized by the Gutzwiller factors
defined in Eq.[\ref{factor}], leading to the values shown in Fig.\ref%
{parameter_fig}. The sign changes in $G^{RVB}(\boldsymbol{k},\omega =0)$
occur on a Luttinger surface of zeroes, which coincides with the umklapp
surface, and on a Fermi pocket of infinities. The parameter $\mu _{p}$ was
adjusted at each $x$ to give the correct area for the hole pockets (see Fig.%
\ref{parameter_fig}). 

\begin{figure}[tbp]
\includegraphics[width=9.0cm,height=12.0cm]{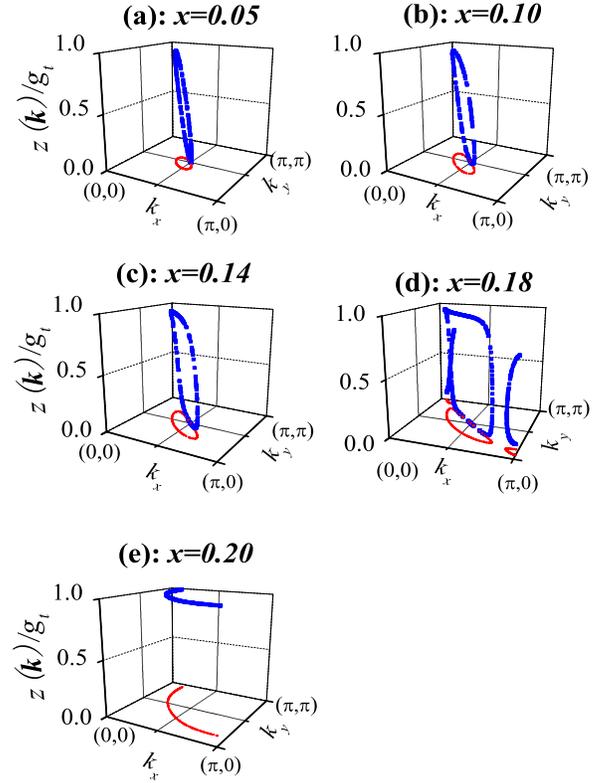}
\caption{(Color online) The spectral weight distribution $z(\boldsymbol{k}%
)/g_{t}$ around the hole quasiparticle pockets in the normal pseudogap
phase. }
\label{pocket_weight_fig}
\end{figure}

As we can see from Fig.\ref{LSR_fig}, the hole pocket evolves gradually into
a more normal surface in panel(e) as $x$ increase. The spectral weight of
the quasiparticle pole varies strongly around the pocket as illustrated in
Fig.\ref{pocket_weight_fig}. In particular it is very small on the outer
edge of the pocket closest to the Luttinger surface of zeroes and vanishes
completely along the nodal directions ((1,1) directions). Along this
direction there is only a single sign change in $G^{RVB}(\boldsymbol{k}%
,\omega =0)$ at the inner pocket edge.

\begin{figure}[tbp]
\includegraphics[width=9.0cm,height=10.0cm]{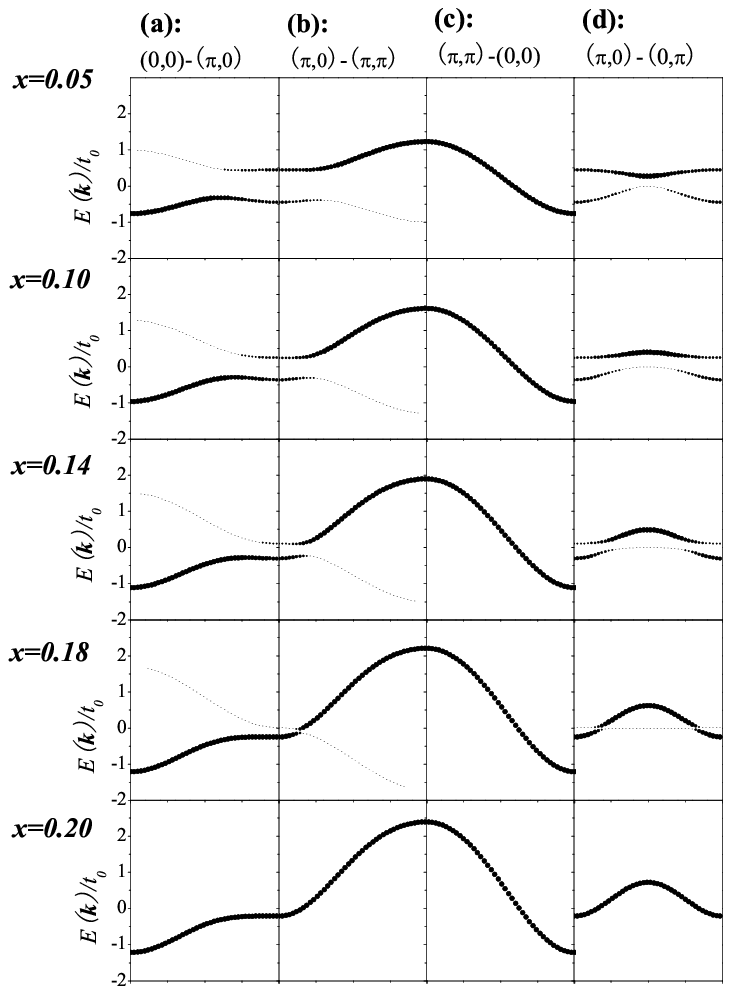}
\caption{(Color online) Quasiparticle dispersion in the normal pseudogap
phase (Eq.[\protect\ref{PG}]) along some high symmetry directions. The
thickness of the lines is proportional to the spectral weight $z(\boldsymbol{%
k})/g_{t}$ of the quasiparticle.}
\label{NS_spectrum_fig}
\end{figure}

In Fig.\ref{NS_spectrum_fig}, the dispersion of the quasiparticle poles $%
E_{i}(\boldsymbol{k})$ in the coherent part of $G^{RVB}$ together with their
spectral weight, $z_{i}(\boldsymbol{k})$, are shown. We see that in general
there are two quasiparticle bands $E_{i}(\boldsymbol{k})$ with strongly
varying spectral weight. Over most of the Brillouin zone the bands are
separated by an energy gap with only the lower band occupied. The
distribution of spectral weight between the two bands is determined by the
proximity to the umklapp surface. Near this surface the weight is equally
divided but away from this surface only a small admixture is induced by the
anomalous self energy $\Sigma _{R}$. Also along the nodal directions $\Sigma
_{R}\rightarrow 0$, so that in these directions the two bands coalesce into
a single band.

\begin{figure}[tbp]
\includegraphics[width=9.5cm,height=8.5cm]{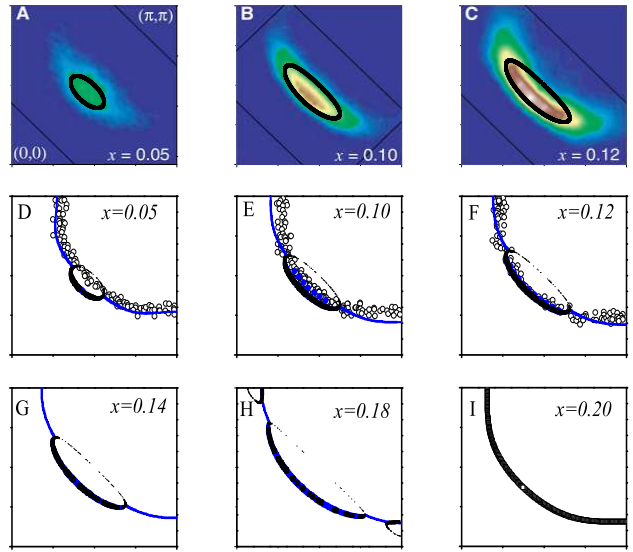}
\caption{(Color online) Comparison between our theory and some recent ARPES
experiments on $Ca_{2-x}NaCuO_{2}Cl$ by K. M. Shen et al. \protect\cite{K.
M. Shen} The experimental results are re-plotted in panels (A-F). Panels
(A-C) show the distribution of spectral weight in the Brillouin zone within
a $\pm 10meV$ window around the Fermi level. The open/solid circles in
panels (D-F) are detected by this experiment to determine the Fermi surface.
The pockets in panel (A-I) show the infinities of $G(\boldsymbol{k},0)$ in
the normal pseudogap phase. The \textquotedblleft Fermi
surfaces\textquotedblright\ (blue and black curves) shown in panels (D-I)
are defined by the minimum distance from the lower quasiparticle band to the
Fermi level in our calculation along radial directions centered at $(\protect%
\pi ,\protect\pi )$. The thickness of the curve in panels (D-I) represents
the spectral weight $z(\boldsymbol{k})/g_{t}$ of the quasiparticle.}
\label{comparison_fig}
\end{figure}

\begin{figure}[tbp]
\includegraphics[width=9cm,height=11.cm]{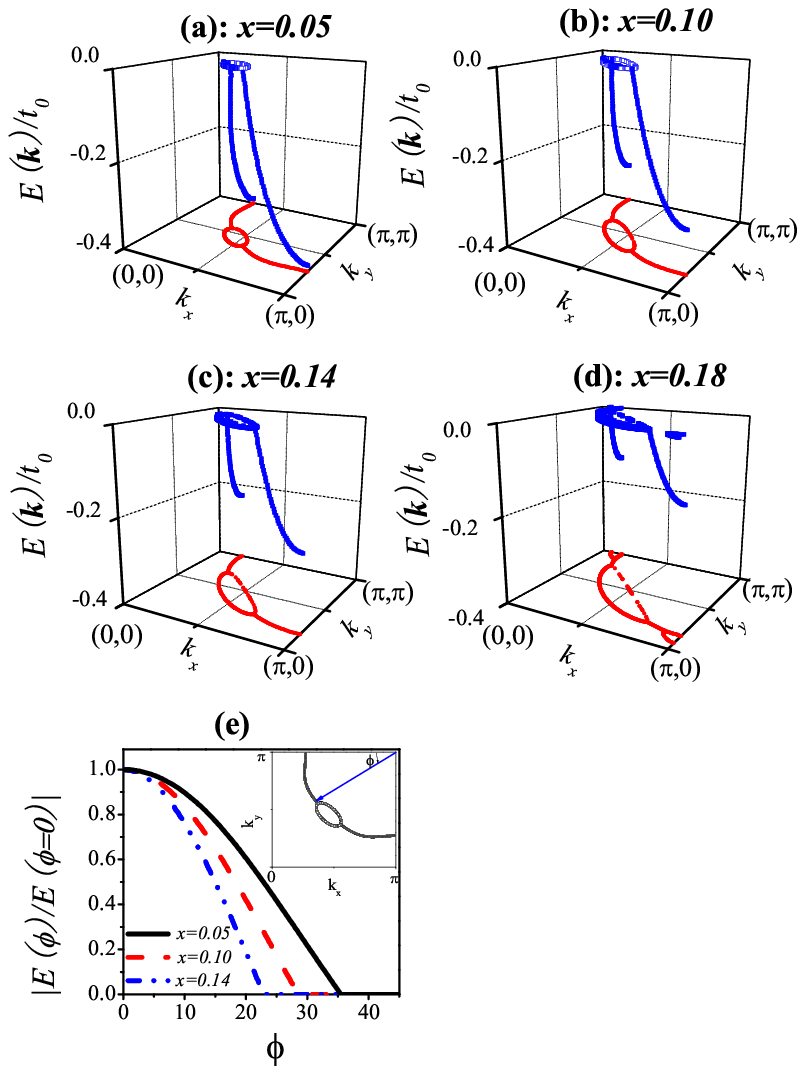}
\caption{(Color online) Panel(a-d) show the dispersion of the lower
quasiparticle band $E(\boldsymbol{k})$ on the \textquotedblleft Fermi
surface\textquotedblright\ and on the hole pockets for various dopings. In
panel(e) we show the angular dependence of $\left\vert E(\boldsymbol{k}%
)\right\vert $ on the \textquotedblleft Fermi surface\textquotedblright . $%
\protect\phi $ is defined in the inset of panel (e).}
\label{FS_dispersion_fig}
\end{figure}

In order to compare to ARPES experiments on $Ca_{2-x}Na_{x}CuO_{2}Cl_{2}$ 
\cite{K. M. Shen}, we prepared a single plot which combines contours of the
hole Fermi pockets and also the minimum energy gap lines in other parts of
zone. These are illustrated in Fig.\ref{comparison_fig} together with the
ARPES results. There is good agreement between the two sets of curves with
the exception of the outer edge of the hole pockets which has not been
reported in the ARPES experiments. The predicted spectral weight of the
quasiparticle band on these outer edges is very small (see Fig.\ref%
{pocket_weight_fig}). Nonetheless it would be important to search more
closely for a weak signal on this edge. In Fig.\ref{FS_dispersion_fig}, we
plot the dispersion of the lower quasiparticle band $E(\boldsymbol{k})$ on
the \textquotedblleft Fermi surface\textquotedblright\ and on the hole
pockets for several dopings. The angular dependence of $|E(\boldsymbol{k})|$
on the \textquotedblleft Fermi surface\textquotedblright\ is shown in panel
(e). The dispersion drops to zero rather sharply as the angle touches the
pocket direction.

\begin{figure}[tbp]
\includegraphics[width=9.0cm,height=10.0cm]{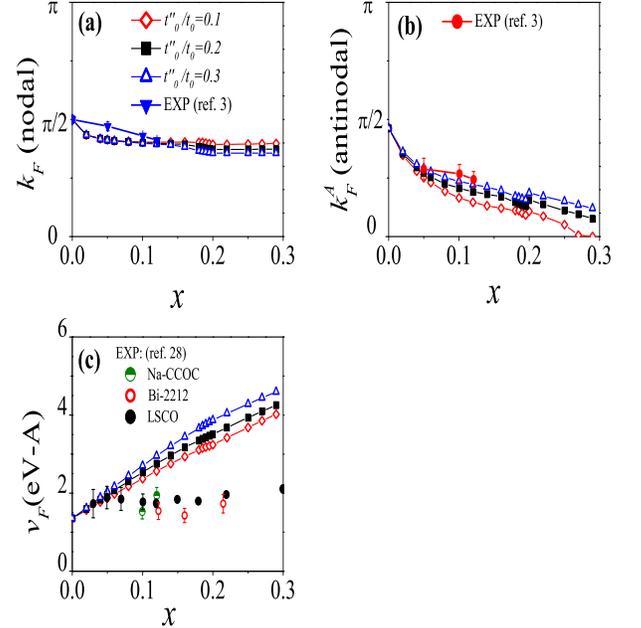}
\caption{(Color online) Nodal Fermi wavevector $k_{F}$, `antinodal Fermi
wavevector' $k_{F}^{A}$, Fermi velocity $v_{F}$ (with values $t_{0}=0.5eV$,
lattice constant $a=4.0\mathring{A}$). The experimental data in panel (a,b)
for nodal and `antinodal Fermi wavevector' are for $%
Ca_{2-x}Na_{x}CuO_{2}Cl_{2}$ \protect\cite{K. M. Shen} while the
experimental results in panel (c) are for $Ca_{2-x}Na_{x}CuO_{2}Cl_{2}$, $%
La_{2-x}Sr_{x}CuO_{4}$, and $Bi_{2}Sr_{2}CaCu_{2}O_{8+\protect\delta }$ 
\protect\cite{X. J. Zhou}.}
\label{kf_vf_fig}
\end{figure}

In Fig.\ref{kf_vf_fig} we show the hole density dependence of the two
features, the Fermi wavevector, $k_{F}$ (with $\boldsymbol{k}_{F}=k_{F}(1,1)$%
) at the hole pocket along the nodal line $(1,1)$ and the intercept of the
minimum energy gap line at the Brillouin zone boundary connecting $(\pi
,0)-(\pi ,\pi )$ antinodal-$k_{F}^{A}$ (with $\boldsymbol{k}%
_{F}^{A}=(k_{F}^{A},\pi )$). Again the agreement is very good with ARPES
experiments on $Ca_{2-x}Na_{x}CuO_{2}Cl_{2}$ \cite{K. M. Shen}. These two
wavevectors have been interpreted as key parameters of an underlying Fermi
surface. But such an interpretation clearly violates the LSR since the
enclosed area would require electron doping in contrast to our ansatz for
the Green's function. As we remarked earlier, our ansatz reconciles the
ARPES measurements and the LSR.

Also in Fig.\ref{kf_vf_fig} we show the Fermi velocity $v_{F}$ along the
nodal direction. This shows rather more variation with the hole doping, $x$,
than the direct calculations from the Gutzwiller projected variational
wavefunction \cite{VF-VMC} but the qualitative behavior is similar. However,
it deviates substantially from experiment with increasing $x$. The
discrepancy of $v_{F}$ at large $x$ could be due to the oversimplified model.

Another quantity of interest is the coherent quasiparticle contribution to
the tunneling density of state (DOS) defined as 
\begin{equation}
N_{T}(\omega )=\sum_{\boldsymbol{k},i=1,2}z_{i}(\boldsymbol{k})\delta
(\omega -E_{i}(\boldsymbol{k}))  \label{DOS}
\end{equation}

\begin{figure}[tbp]
\includegraphics[width=9.0cm,height=7.0cm]{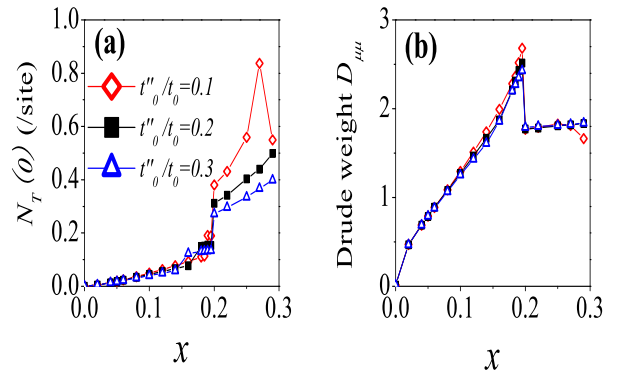}
\caption{(Color online) Density of states at the Fermi level $N_{T}(0)$ and
the Drude weight $D_{\protect\mu \protect\mu }$ in the normal pseudogap
phase.}
\label{DOS_Drude_fig}
\end{figure}

\begin{figure}[tbp]
\includegraphics[width=9.0cm,height=12.0cm]{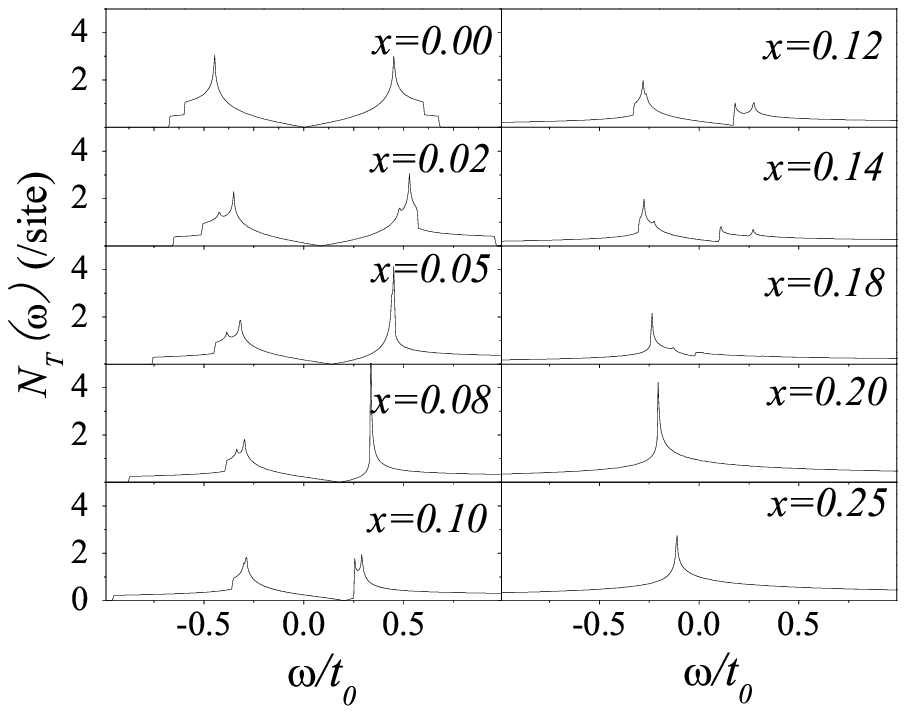}
\caption{(Color online) Density of states $N_{T}(\protect\omega )$ in the
normal pseudogap phase. The zero energy values $N_{T}(0)$ are shown in Fig.%
\protect\ref{DOS_Drude_fig}(a). }
\label{DOS_fig}
\end{figure}
The value at the chemical potential, $N_{T}(0)$ is shown for a series of
hole densities in Fig.\ref{DOS_Drude_fig}(a). $N_{T}(0)$ rises linearly as
the hole density $x$ is increased from zero in the pseudogap phase. This
behavior is similar to the Drude weight, $D_{\mu \upsilon }$, calculated by
integrating around the Fermi surfaces of the hole pockets%
\begin{equation}
D_{\mu \upsilon }=\frac{2}{(2\pi )^{2}}\frac{e^{2}}{\hbar }\int \frac{v_{\mu
}v_{\upsilon }}{\left\vert \boldsymbol{v}\right\vert }dS_{F}
\label{Drude_weight}
\end{equation}%
In the square lattice $D_{\mu \upsilon }$ is a diagonal tensor with element $%
D_{\mu \mu }$ shown in Fig.\ref{DOS_Drude_fig}(b).

The density dependence of the RVB gap parameter, $\Delta _{0}(x)$ was taken
from the RMF results of Zhang et al.\cite{RMF} This gives a linear drop in $%
\Delta _{0}$ with $\Delta _{0}(x)\rightarrow 0$, as $x\rightarrow x_{c}$
(see in Fig.\ref{parameter_fig}(e)). We have chosen a value of $x_{c}=0.2$.
This linear drop in $\Delta _{0}(x)$ as $x$ increases is in line with ARPES
experiments as emphasized in the recent review by Anderson et al. \cite%
{vanillar} The vanishing of $\Delta _{0}(x)$ at $x=x_{c}$ causes a change in
the form of $G(\boldsymbol{k},\omega )$ with a transition from the doped
spin liquid RVB character to that of a standard Landau Fermi liquid. Then
contours of sign changes in $G(\boldsymbol{k},0)$ that enter the LSR change
their topology and the Luttinger surface of zeroes disappears. Only a
standard closed Fermi surface of infinities exists at $x>x_{c}$.

The detailed form of the transition from a doped RVB spin liquid to a Landau
Fermi liquid depends on the band parameters. For the choice that we made to
correspond to $Ca_{2-x}Na_{x}CuO_{2}Cl_{2}$, there is an additional
topological change in the LSR contours as $x\rightarrow x_{c}$. As
illustrated in Fig.\ref{LSR_fig}, for a narrow range of $x\lesssim x_{c}$ a
new set of electron Fermi pockets appear close to the saddle points, outside
the Luttinger surface of zeroes. These new electron pockets merge with the
hole pockets inside the Luttinger surface of zeroes at $x=x_{c}$ to give a
Fermi surface that crosses the umklapp surface at $x>x_{c}$. In our
parameter choice the saddle points in the band structure are occupied for a
range of hole densities for $x\gtrsim x_{c}$ which requires that the Fermi
surface cross the umklapp surface in this density range. For this parameter
choice, there are two closely spaced topological changes in the LSR
contours. The first is a Lifschitz-type transition in which the shape of the
Fermi surface of infinities undergoes a topological change. The second at $%
x=x_{c}$ is a quantum critical point (QCP) associated with opening of a
single particle gap at the onset of the RVB spin liquid for $x<x_{c}$
leading to a Luttinger surface of zeroes coinciding with the umklapp surface.

These two transitions cause singularities in the density dependence of the
DOS and the Drude weight as illustrated in Fig.\ref{DOS_Drude_fig}. The
strongest singularity appears at the QCP at $x=x_{c}$. A substantial jump
appears in the DOS at $x=x_{c}$. The DOS continues to rise as $x$ increases
beyond $x_{c}$ which is associated with the approach of the Fermi energy to
the van Hove singularity at the saddle points in the band structure. A
divergence of the DOS at the Fermi energy at $x>x_{c}$ has not been observed
to our knowledge but this might be due to different band parameters or
possibly to a suppression of the divergence due to impurity scattering. The
full energy dependence of the DOS is shown in Fig.\ref{DOS_fig}.

The QCP in our phenomenological theory is qualitatively similar to that
inferred by Loram,\ Tallon and collaborator from experiments \cite{Loram,
Tallon}. Their analysis emphasized the singularities in thermodynamic
quantities such as specific heat, magnetic susceptibility etc. Our
phenomenological theory is restricted at present to zero temperature and so
is not suitable for detailed comparison. However, we note that the main
conclusion they drew, that a partial gap opened up at $x<x_{c}$, is in
agreement with our phenomenological theory.

\section{Electronic Properties of the Superconducting State}

We turn now to the evolution of the electronic properties when the system
enters a d-wave superconducting state. Superconductivity is introduced by
the addition of the new term in the self-energy which follows from the
standard Green's function theory of a superconductor \cite{AGD}. Solving the
coupled equations which connect the regular ($G^{S}(\boldsymbol{k},\omega )$%
) and anomalous ($F(\boldsymbol{k},\omega )$) Green's functions (see Fig.\ref%
{pseudogap_fig})

\begin{figure}[tbp]
\includegraphics[width=9.0cm,height=5.0cm]{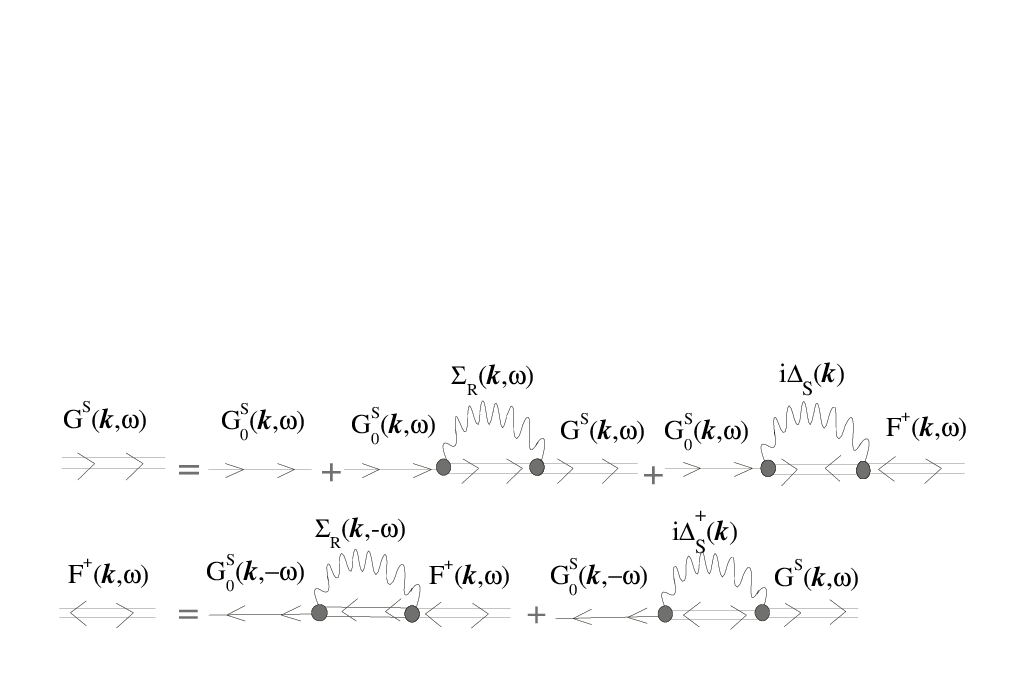}
\caption{(Color online) Regular and anomalous Green's functions $G(%
\boldsymbol{k},\protect\omega )$, $F^{\dagger }(\boldsymbol{k},\protect%
\omega )$ which appear in the coupled Eq.[\protect\ref{coupled1},\protect\ref%
{coupled2}].}
\label{pseudogap_fig}
\end{figure}

\begin{eqnarray}
&&(\omega -\xi (\boldsymbol{k})-\Sigma _{R}(\boldsymbol{k},\omega ))G^{S}(%
\boldsymbol{k},\omega )  \notag \\
&&-i\Delta _{S}(\boldsymbol{k})F^{\dagger }(\boldsymbol{k},\omega )=1
\label{coupled1}
\end{eqnarray}%
\begin{eqnarray}
&&(-\omega -\xi (\boldsymbol{k})-\Sigma _{R}(\boldsymbol{k},-\omega
))F^{\dagger }(\boldsymbol{k},\omega )  \notag \\
&&-i\Delta _{S}^{\dagger }(\boldsymbol{k})G^{S}(\boldsymbol{k},\omega )=0
\label{coupled2}
\end{eqnarray}%
leads the result quoted earlier in Eq.[\ref{SC}]%
\begin{eqnarray}
&&[\omega -\xi (\boldsymbol{k})-\Sigma _{R}(\boldsymbol{k},\omega )  \notag
\\
&&-\frac{\left\vert \Delta _{S}(\boldsymbol{k})\right\vert ^{2}}{(\omega
+\xi (\boldsymbol{k})+\Sigma _{R}(\boldsymbol{k},-\omega ))}]G^{S}(%
\boldsymbol{k},\omega )=1  \label{SC2}
\end{eqnarray}%
The gap function is assumed to have a d-wave form and it is related to the
anomalous Green's function 
\begin{equation}
\Delta _{S}(\boldsymbol{k})=\int d^{2}\boldsymbol{k}^{\prime }d\omega g(%
\boldsymbol{k-k}^{\prime })F(\boldsymbol{k}^{\prime },\omega )  \label{gap}
\end{equation}%
where $g(\boldsymbol{k-k}^{\prime })$ is the d-wave pairing interaction.

\begin{figure}[tbp]
\includegraphics[width=7.0cm,height=6.0cm]{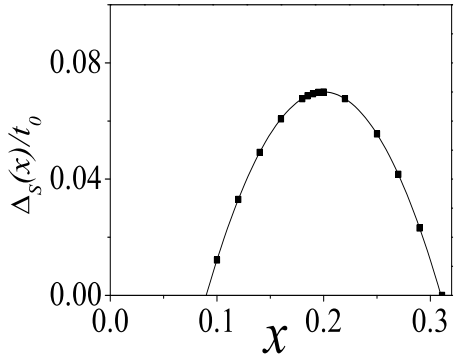}
\caption{(Color online) The phenomenological form of the superconducting gap
that follows from a parabolic relation for $T_{c}(x)$: $\Delta
_{S}(x)/t_{0}=0.07(1-82.6(x-0.2)^{2})$.}
\label{deltasc_fig}
\end{figure}

The strength of the superconductivity is determined by the magnitude of the
gap squared. We assume a simple parabolic form for $T_{c}(x)$ \cite{M. R.
Presland} to mimic experiment (see Fig.\ref{deltasc_fig}) and the gap is
scaled with the superconducting, $T_{c}$. 

We begin by examining the form of the LSR in the superconducting state. By
inspecting the function $G_{coh}^{S}(\boldsymbol{k},0)$ defined in Eq.[\ref%
{SC}] we see at once that the denominator diverges on the contours defined
by $\xi _{0}(\boldsymbol{k})=0$ and by $\xi (\boldsymbol{k})+\Sigma _{R}(%
\boldsymbol{k},0)=0$. These are the same LSR contours that occur in the
normal state. The only difference is that the second set of contours now
define zeroes of $G_{coh}^{S}(\boldsymbol{k},0)$ not infinities. The Fermi
surface of the hole pockets is gapped except along the nodal line where both 
$\Delta _{R}(\boldsymbol{k})=\Delta _{S}(\boldsymbol{k})=0$. The form (\ref%
{SC}) for $G_{coh}^{S}(\boldsymbol{k},\omega )$ then continues to satisfy
the LSR.

\begin{figure}[tbp]
\includegraphics[width=9.0cm,height=8.0cm]{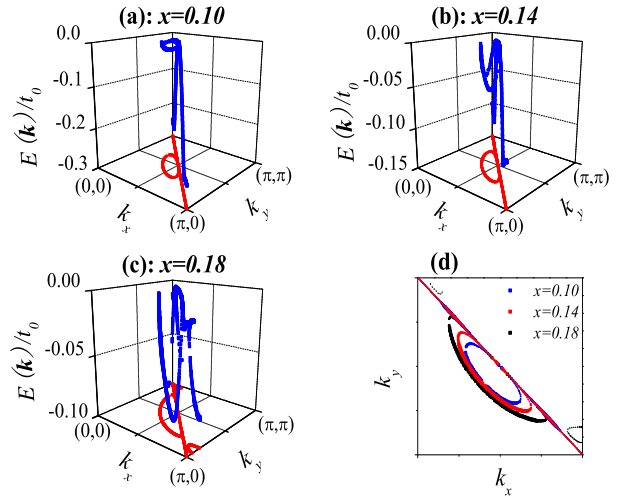}
\caption{(Color online) The dispersion (panels (a-c)) and weight (panel (d))
on the Luttinger surface with nonzero superconducting gap. There are 2 bands
below the Fermi level, and shown here is the one closer to Fermi level.
Around $(\protect\pi ,0)$ and $(0,\protect\pi )$, there is a substantial
part of the spectral weight located at the lower band, only a small part
remains on the band closer to the Fermi level as indicated in Fig\protect\ref%
{spectrum_SC_fig}.}
\label{LS_SC_fig}
\end{figure}

\begin{figure}[tbp]
\includegraphics[width=9.0cm,height=12.0cm]{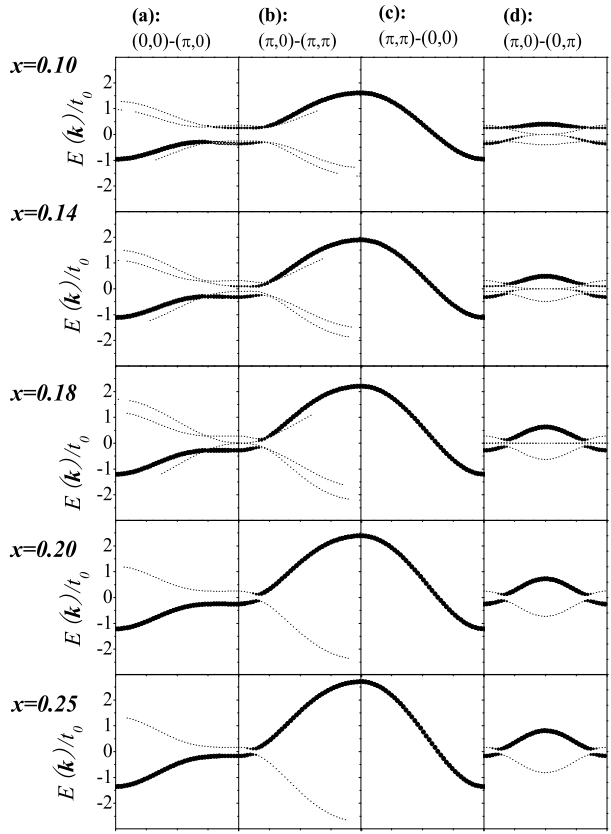}
\caption{(Color online) Quasiparticle dispersion $E(\boldsymbol{k})$ in the
superconducting state (Eq.[\protect\ref{SC}]) along high symmetry
directions. The thickness of the lines is proportional to the spectral
weight $z(\boldsymbol{k})/g_{t}$.}
\label{spectrum_SC_fig}
\end{figure}

In the superconducting state the quasiparticle poles are given by solutions
to a quartic equation%
\begin{eqnarray}
&&(\omega ^{2}-\xi ^{2}(\boldsymbol{k}))(-\omega ^{2}+\xi _{0}^{2}(%
\boldsymbol{k}))+2\Delta _{R}^{2}(\boldsymbol{k})(\omega ^{2}-\xi _{0}(%
\boldsymbol{k})\xi (\boldsymbol{k}))  \notag \\
&&+\Delta _{S}^{2}(\boldsymbol{k})(\omega ^{2}-\xi _{0}^{2}(\boldsymbol{k}%
))-\Delta _{R}^{4}(\boldsymbol{k})=0  \label{SC_poles}
\end{eqnarray}%
which results in a further splitting of the quasiparticle bands. These are
illustrated in Fig.\ref{spectrum_SC_fig} for a number of hole densities. The
spectral weight redistribution is small when the original quasiparticle
energies are away from the chemical potential. This can be seen for example
in panel (d) of Fig.\ref{spectrum_SC_fig} which shows the quasiparticle
bands and their weight along the umklapp surface connecting $(\pi ,0)-(0,\pi
)$. Here we see comparable weights only when the hole Fermi pockets are
nearby.

\begin{figure}[tbp]
\includegraphics[width=9.0cm,height=12.0cm]{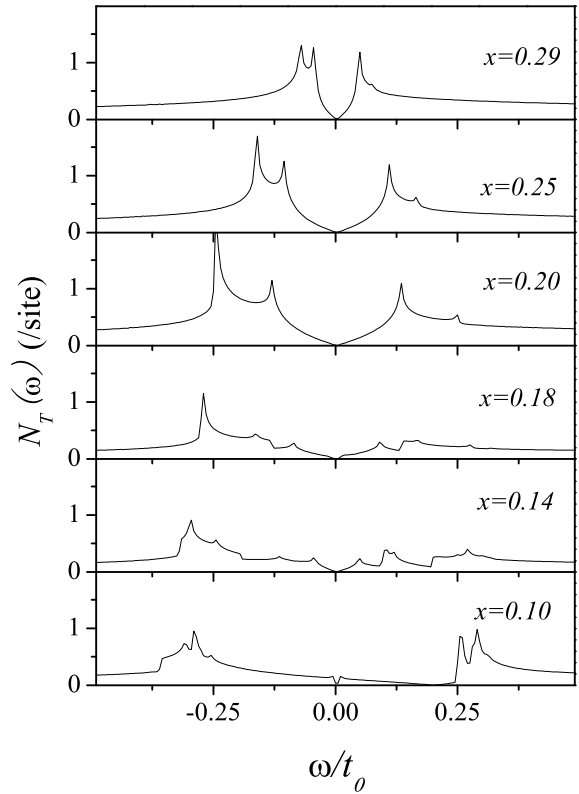}
\caption{(Color online) Density of states $N_{T}(\protect\omega )$ in the
presence of a nonzero superconducting gap. The additional peaks closest to
Fermi level are determined by the superconducting gap which opens on the
hole pockets for $x<x_{c}$. For $x\geq x_{c}$ the peaks come from the
antinodal directions for $(\pm \protect\pi ,0)$, $(0,\pm \protect\pi )$.}
\label{DOS_SC_fig}
\end{figure}

The superconducting energy gap modifies the DOS, $N_{T}(\omega )$. As
illustrated in Fig.\ref{DOS_SC_fig} the opening of the superconducting gap
along the Fermi pockets (shown in Fig.\ref{LS_SC_fig}) leads to a pseudogap
in $N_{T}(\omega )$ accompanied by van Hove singularities at the gap maxima.
Note these van Hove singularities do not occur at the usual antinodal $%
\boldsymbol{k}$-points $(\pm \pi ,0)$, $(0,\pm \pi )$ but rather on the
extremities of the hole Fermi pockets. This is because the form Eq.[\ref{SC}%
] we chose for the superconducting state does not represent a merging of the
two gaps $\Delta _{R}(\boldsymbol{k})$ and $\Delta _{S}(\boldsymbol{k})$,
rather both gaps keep their own identity. Further consideration of this
point would require a better microscopic understanding of the coexistence of
d-wave superconductivity with the RVB spin liquid correlations than our
simple phenomenological ansatz.

Another clear feature in $N_{T}(\omega )$ is the van Hove singularity
associated with antinodal saddle points in the RVB quasiparticle bands at
negative energy. This feature does not appear in the STM experiments. This
could be due to the strong local variations in the hole density.

\section{Discussion and Conclusion}

The ansatz for the Green's function Eq.[\ref{PG}] is motivated in the first
place by the form of the Green's function derived recently by Konik and
coworkers \cite{Konick, KRT} for a doped spin liquid composed of an array of
2-leg Hubbard ladders. A second important input is the renormalized mean
field derived many years ago by Zhang and coworkers \cite{RMF}. As
emphasized recently by Andseron and coworkers \cite{vanillar}, the RMF,
although it treats the strong correlations simply by Gutzwiller
renormalization factors, agrees qualitatively and even at times
quantitatively with the results of variational Monte Carlo (VMC)
calculations using Gutzwiller projected wavefunctions \cite{VF-VMC}. The RMF
and VMC calculations at finite doping rely on a broken symmetry to introduce
a gap at the Fermi energy. However, in the underdoped region the gap at the
antinodal regions ($\sim J$) is much larger than the expected
superconducting transition temperature $T_{c}$ as estimated for example by
Wen and Lee \cite{Wen and Lee}. So one should expect this gap to persist in
the normal state at temperatures $T\geq T_{c}$, and so should be a property
of the RVB spin liquid and not related to a broken symmetry. Another key
feature of the underdoped region is that the Drude weight scales as the hole
concentration. As a result the simplest explanation for these two properties
is that the Fermi surface is partially truncated in a doped RVB spin liquid
rather like the partial truncation that occurs through spin density waves in 
$Cr$ and its alloys before the commensurate antiferromagnetic state is
reached \cite{Cr and alloys}. Here the key difference to the $Cr$ alloys is
the absence of a broken symmetry and long range order. Theses properties are
reconciled by our ansatz for the normal state. In the superconducting state
we introduced $\Delta _{S}$ and $\Delta _{R}$ as separate gaps whereas in
the RMF and VMC calculations there is only a single gap function opening up
along a single Fermi surface. So it would appear that our ansatz is
qualitatively different to the RMF and VMC theories in the superconducing
state. However, one cannot be sure of this since the key quantity is the
zero frequency Green's function $G(\boldsymbol{k},0)$ which is not directly
available from a variational wavefunction. We note that there have been
recent reports on the numerical study of the pseudogap phase by using
cluster perturbation theory (CPT) \cite{D. Senechal} and by the dynamical
mean field theory (DMFT) \cite{A. Macridin} of the 2-Dimension Hubbard model.

Our ansatz does not deal with the origin of the d-wave superconductivity 
\textit{per se}. The relevant issue is the stability of the normal state as
a doped RVB state. This issue was addressed in the case of the doped array
of 2-leg Hubbard ladders by KRT \cite{KRT}. They showed that strong residual
interactions acting on the hole pockets lead to a d-wave superconducting
state or possibly a spin density wave state depending on the parameters.
There is strong reason to believe that similar effects should occur here in
the 2-dimensional doped RVB spin liquid. Indeed such effect were conjectured
by Honerkamp \cite{Honerkamp}, Laeuchli and coworkers \cite{Laeuchli} in
their analysis of the functional RG calculation for the 2-dimensional $%
t-t^{\prime }-U$ Hubbard model. On a qualitative level we know that the RVB
spin liquid has enhanced response in both d-wave pairing and
antiferromagnetic channels so that the Fermi pockets are moving in a
background which is highly polarizable in both these channels which can lead
to corresponding instabilities. The challenge is to develop a proper
microscopic theory to describe the competition between the two
instabilities. Note unlike the case of heavy fermions where a
superconducting dome straddles the QCP associated with onset of long range
AF order \cite{N. D. Mathur}, here superconductivity seems to be most stable
near the onset of the RVB phase with short range correlations. The good
agreement that this ansatz displays with many properties of the anomalous
normal state in the pseudogap region is evidence that it contains at least
elements of the correct physics.

The authors acknowledge the support from Centre of Theoretical and
Computational Physics and Visiting Professorship at The University of Hong
Kong. T. M. Rice also acknowledges support at ETH from the MANEP program of
the Swiss National Foundation. 

\end{document}